\documentclass[%
 reprint,
superscriptaddress,
 amsmath,amssymb,
 aps,
]{revtex4-2}

\usepackage{graphicx}
\usepackage{dcolumn}
\usepackage{bm}

\begin{document}

\preprint{APS/123-QED}

\title{Large Gap Quantum Anomalous Hall Effect in a Type-I Heterostructure Between a Magnetically Doped Topological Insulator and Antiferromagnetic Insulator }

\author{Anh Pham}
 \affiliation{Center for Nanophase Materials Sciences, Oak Ridge National Laboratory, Oak Ridge, Tennessee 37831, USA}
\author{Ling-Jie Zhou} 
 \affiliation{Department of Physics, The Pennsylvania State University, University Park, Pennsylvania 16802, USA}
\author{Yi-Fan Zhao}
 \affiliation{Department of Physics, The Pennsylvania State University, University Park, Pennsylvania 16802, USA}
\author{Cui-Zu Chang}
 \affiliation{Department of Physics, The Pennsylvania State University, University Park, Pennsylvania 16802, USA}
\author{Timothy Charlton} 
 \affiliation{Spallation Neutron Source, Oak Ridge National Laboratory, Oak Ridge, Tennessee 37831, USA}
\author{P. Ganesh}
 \email{ganeshp@ornl.gov}
\affiliation{Center for Nanophase Materials Sciences, Oak Ridge National Laboratory, Oak Ridge, Tennessee 37831, USA}

\date{\today}

\begin{abstract}
Heterostructures between topological insulators (TI) and magnetic insulators represent a pathway to realize the quantum anomalous Hall effect (QAHE). But major challenges remain, typical heterostructures have weak magnetic exchange interactions across the interface, resulting in small inverted band gaps. Magnetic doping can further increase the band gap due to the additional Zeeman effect, but gaps remain small at low concentrations of magnetic dopants, with higher concentrations resulting in shifting of the Fermi-level, that can render the heterostructure metallic. In addition, chemical bonding across interfaces can lead to imperfect interfaces that might induce metallicity or further lower the gap size. Using density functional theory based systematic screening and investigation of thermodynamic, magnetic and topological properties of heterostructures, we demonstrate that forming a type-I heterostructure between a wide gap antiferromagnetic insulator Cr$_2$O$_3$ and a TI-film, such as Sb$_2$Te$_3$, can lead to pinning of the Fermi-level at the center of the gap, even when magnetically doped. Cr-doping in the heterostructure increases the gap to $\sim$ 64.5 meV, with a large Zeeman energy from the interfacial Cr dopants, thus overcoming potential metallicity due to band bending effects.  By fitting the band-structure around the Fermi-level to a 4-band {\sl k.p} model Hamiltonian, we show that Cr doped Sb$_2$Te$_3$/Cr$_2$O$_3$ is a Chern insulator with a Chern number C = -1. Transport calculations further show chiral edge-modes localized at the top/bottom of the TI-film to be the dominant current carriers in the material.  Our predictions of a large interfacial magnetism due to Cr-dopants, that coupled antiferromagnetically to the AFM substrate is confirmed by our polarised neutron reflectometry measurements on MBE grown Cr doped Sb$_2$Te$_3$/Cr$_2$O$_3$ heterostructures, and is consistent with a positive exchange bias measured in such systems recently. Consequently, Cr doped Sb$_2$Te$_3$/Cr$_2$O$_3$ heterostructure represents a promising platform for the development of functional topological magnetic devices, with high tunability. 
\end{abstract}

\maketitle


\section{\label{sec:level1} Introduction}

The quantum anomalous Hall effect is characterized by a chiral edge state in an insulating topological system with broken time-reversal symmetry.  These edge states carry a dissipationless current with zero longitudinal resistance and quantized Hall resistance [1-3]. As a result, the QAHE has been proposed to have a broad range of applications in electronic and spintronic devices as well as topological quantum computing [4]. To realize the QAHE, it is expected that a material system has strong spin-orbit coupling (SOC) and magnetic exchange energy to induce nontrivial topology in the bulk band structure [1-3]. Furthermore, due to the breaking of time reversal symmetry an exchange gap can be opened on the surface of the topological insulator(TI). Consequently, early theoretical studies suggested that the TI thin films doped with transition metal elements in a diluted concentration can be used a platform for the realization of the QAHE. In theory, the QAHE appears in TI thin films with Cr or Fe doping, but it is absent in TI thin films with V or Ti doping. The absence of the QAHE in the latter two systems is due to the formation of the impurity states located at the magentic exchange gap, which hinders the appearance of the insulating state [5,6]. This prediction was subsequently confirmed in experimental studies of doped Bi(Sb)-based chalcogenides thin films with V and Cr dopants [7-13]. However, the observing temperature of the QAHE in magnetically doped topological insulator films/heterostructures has been less than 1K thus limiting its functional applications. \\
\indent The low observing temperature of the QAHE in magnetically doped TI films/heterostructures has been attributed to the inhomogeneity of magnetic dopants, i.e. chemical potential fluctuation. This inhomogeneity can break the inversion symmetry in the TI thin film, resulting in metallic bulk bands [14]. In addition, the inhomogeneity can also reduce the effective magnetic exchange energy gap since the gap opening at the Dirac point due to time reversal symmetry breaking and the magnetic interaction between TM dopants are strongly dependent on the spatial distribution of magnetic defects [15]. Another draw back of conventional doping technique to realize the QAHE is that only diluted doping is allowed so that the Fermi level lies inside the bulk gap, while high doping is not favourable because it can lead to the weakening of spin-orbit coupling that will eventually destroy the nontrival surface states of the TI thin films [16,17], even though the high doping level of transition metal dopant can enhance the Curie temperature to achieve a larger magnetic exchange gap. To overcome this fundamental material limitation, the modulation doping technique [18] has been demonstrated to increase the observing temperature of the QAHE by increasing the magnetic doping concentration on the two surface layers of the magnetically doped sandwich heterostruture samples [19,20].  \\
\indent In addition to the magnetic doping, the QAHE has also been proposed to be realized in different heterostructure configurations between a TI and a ferromagnetic/antiferromagnetic insulator. Different ferromagnetic insulating materials such as EuS [21,22], GdN [23], BaFe$_{12}$O$_{19}$ [24], Cr$_2$Ge$_2$Te$_6$ [25], ferrimagnet iron garnets (YIG/TIG) [26,27], and CrI$_3$ [28,29] have all demonstrated their ability to induce the QAHE on the surface of a TI. In these heterostructures, the magnetic exchange is induced on the TI's surface via magnetic proximity effect. However, the induced moment tends to be small and the effect can be short-range thus limiting the value of the gap opening due to time-reversal symmetry breaking. Antiferromagnetic (AFM) insulators with uncompensated intralayer interactions such as Cr$_2$O$_3$ [30] and LaCoO$_3$ [31] have also been reported to induce the proximity induced magnetic order when interfacing with a TI thin film. AFM insulating substrates have several notable advantages, they lack stray dipole fields and are robust against external perturbing fields. Further, AFM insulators such as Cr$_2$O$_3$ also show ultrafast dynamics [32]. In addition, since the Néel temperature in many AFM insulators is much higher than the Curie temperature in FM insulators, this can enhance the Curie temperature of the proximity induced FM order in the TI [30]. \\ 
\indent Since Cr$_2$O$_3$ is a wide band gap AFM insulator with a high Néel temperature (T$_N$ $\sim$307 K) [30], we systematically examine the effect of Cr doped TI in the heterostructure configuration with TI using first principles calculation. Experimentally many Cr$_2$O$_3$ thin films tend to be fabricated on substrate like Al$_2$O$_3$ [33], we set our in-plane lattice constant to the value of Al$_2$O$_3$ (4.756 \r{A}) [34] similar to our recent experimental fabrication of Cr doped Sb$_2$Te$_3$/Cr$_2$O$_3$ on Al$_2$O$_3$ substrates [30]. We screened several TI materials to assess lattice mismatch with the AFM insulator. As shown in Table 1, both Sb$_2$Te$_3$ and Bi$_2$Se$_3$ have the lowest lattice mismatch with the Cr$_2$O$_3$ substrate. But we find that Bi$_2$Se$_3$ and Cr$_2$O$_3$ interface to be highly reactive, with strong Se-Cr bonding, resulting in a metallic ground state, making it unsuitable for the realization of the QAHE. Screening over magnetic dopants in tellurides suggests magnetic Cr-atoms to be ideal substitutional dopants at the metal sites [35] similar to our earlier studies on SnTe [36]. We hence perform computations for Cr doped Sb$_2$Te$_3$, due to low lattice mismatch, and favourable defect properties. It forms an ideal interface with Cr$_2$O$_3$ with the Fermi level pinned inside the bulk gap due to a type-I band alignment between Cr$_2$O$_3$ and Cr doped Sb$_2$Te$_3$. We also explore the possibility and effect of modulation doping. We find the gap opening in the bulk is strongly enhanced via modulation doping of Cr dopants at high concentrations at the interface with Cr$_2$O$_3$ with band gaps as large as 64.5 meV,  exhibiting a nonzero Chern number of C = -1. As a result, the modulation doped Sb$_2$Te$_3$:Cr/Cr$_2$O$_3$ heterostructure represents a promising platform to realize a QAHE with potentially high temperature functionality. Interfacing such a heterostructure with an {\sl s}-wave superconductor should lead to a large-gap topological superconductivity [37]. \\
\begin{table}[h!]
\centering
\caption{Lattice constants and strain of different TI layer}
\label{table:I}
\begin{tabular}{c c c c}
 \hline\hline
TI & Lattice(\r{A}) & d(\r{A})$^{\ast}$ & Mismatch($\%$)\\
 \hline
 Sb$_2$Te$_3$ & 4.25 & 2.603 & -1.308 \\ 
 Bi$_2$Se$_3$ & 4.138 & 2.404 & -0.463 \\
 Bi$_2$Te$_3$ & 4.383 & N/A & -2.19 \\ 
\hline
\end{tabular}
\begin{flushleft}*Distance between the 4 quintuple layer TI thin films and Cr$_2$O$_3$\end{flushleft}
\end{table}
\begin{figure}[t]
  \centering
  \includegraphics[width=0.5\textwidth]{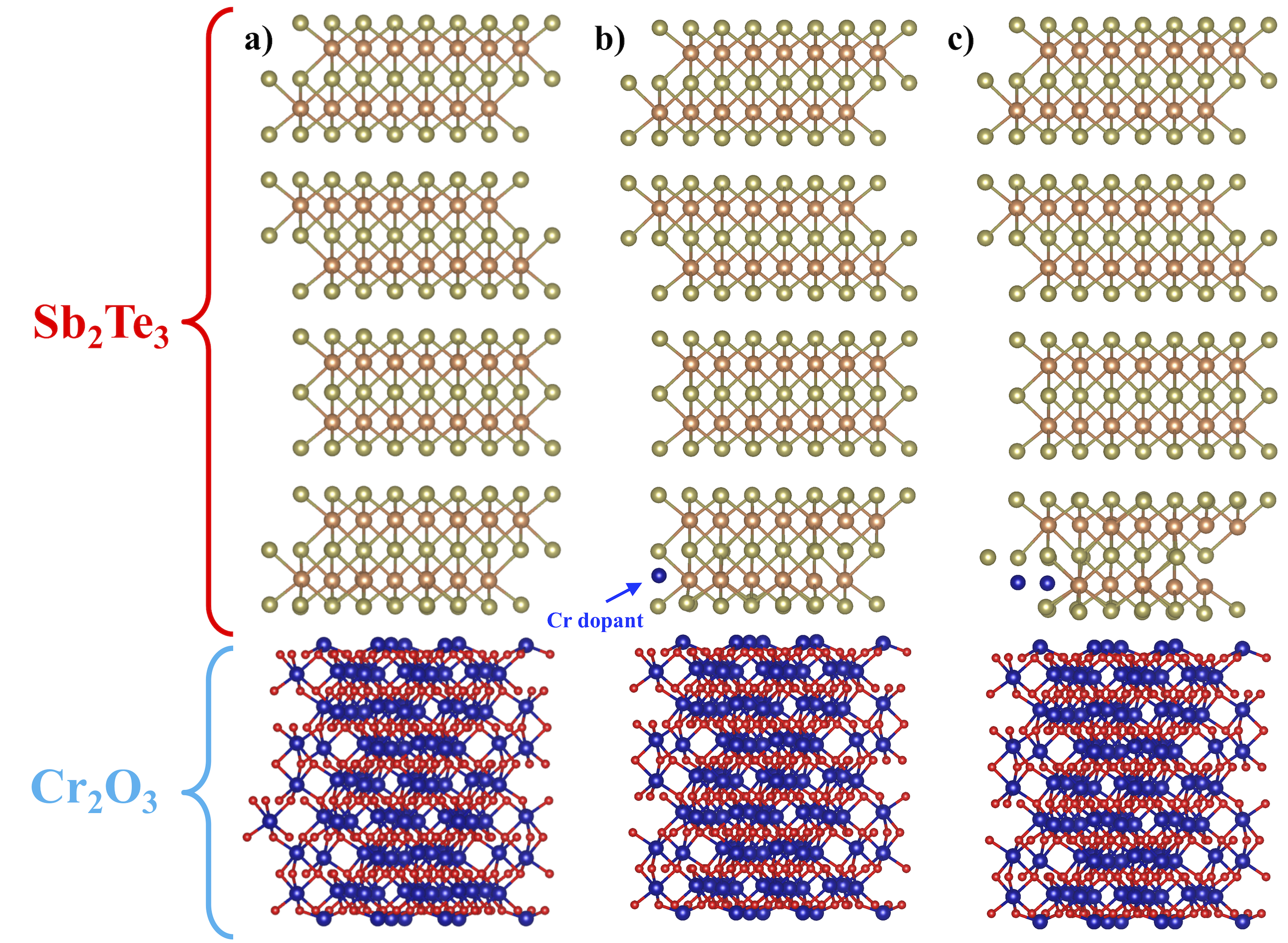}
  \caption {Crystal structure of Sb$_2$Te$_3$/Cr$_2$O$_3$ heterostructure. a) Pristine heterostructure. b) Structure with one Cr dopant substituting Sb at the interface. c) Structure with two Cr dopants substituting Sb at the interface.} 
\end{figure}
\section{\label{sec:level2}Experimental and computational methods}
\textit{Density functional calculation (DFT)}: The calculations were done using the Perdew-Burke-Ernzerhof (PBE) [38] functional with projected augmented wave (PAW) [39] method implemented in the VASP code. An effective Hubbard-U value [40] of U = 4.0 eV and J = 0.58 eV [41] as tested in previous study of Cr$_2$O$_3$ substrate is added to the Cr-${\sl d}$ orbitals to account for the correlation effects. To simulate the Cr$_2$O$_3$ substrate, a symmetric slab consisting of 24 layers of O and 16 layers of Cr was used with Cr-termination since it was determined to be the most stable structure in a previous theoretical study [41]. We consider 4 quintuple layers (QL) for Sb$_2$Te$_3$ and Bi$_2$Se$_3$ TI thin-films. The in-plane lattice of the heterostructure is set to the experimental lattice of Al$_2$O$_3$ (a = b = 4.756 \r{A}) [34]. To minimize the strain between the TI films and the Cr$_2$O$_3$ layers, different supercell configurations of AFM insulator and TI films were constructed as shown in Table 1. Specifically,  the smallest minimally strained Sb$_2$Te$_3$/Cr$_2$O$_3$ heterostructure consists of a 3 $\times$ 3 $\times$ 1 supercell of Sb$_2$Te$_3$ and a $\sqrt{7}$ $\times$ $\sqrt{7}$ $\times$ 1 supercell of Cr$_2$O$_3$ resulting in a lattice mismatch of 1.3$\%$ for the TI layer. The Bi$_2$Se$_3$/Cr$_2$O$_3$ heterostructure consists of a 2 $\times$ 2 $\times$ 1 supercell of Bi$_2$Se$_3$ and a $\sqrt{5}$ $\times$ $\sqrt{5}$ $\times$ 1 supercell of Cr$_2$O$_3$ with a lattice mismatch of 0.46$\%$. While the Bi$_2$Te$_3$/Cr$_2$O$_3$ heterostructure consists of a 4 $\times$ 4 $\times$ 1 of Bi$_2$Te$_3$ and $\sqrt{13}$ $\times$ $\sqrt{13}$ $\times$ 1 of Cr$_2$O$_3$ with a lattice mismatch of 2.19$\%$. We focus our study primarily on the Sb$_2$Te$_3$/Cr$_2$O$_3$  and Bi$_2$Se$_3$/Cr$_2$O$_3$ since they contain the lowest lattice mismatch. The distance between the TI thin film and the Cr$_2$O$_3$ layers was optimized as shown in Table 1 by determining the minimal energy configuration as the distance between the TI and the AFM layer was varied. An energy cutoff of 400 eV and 320 eV was used for the Bi$_2$Se$_3$ and Sb$_2$Te$_3$ heterostructures respectively. A gamma-center {\bf k} point of 2 $\times$ 2 $\times$ 1 and 1 $\times$ 1 $\times$ 1 was used to optimize the heterostructures of Bi$_2$Se$_3$ and Sb$_2$Te$_3$ respectively. To minimize the interaction  along the z-direction, a vacuum of 20 \r{A} was used for all the structures.\\
\indent \textit{Synthesis of Cr doped Sb$_2$Te$_3$/Cr$_2$O$_3$ hetero-structures}: The Cr$_2$O$_3$ films were grown on heat-treated sapphire (0001) substrates in a pulsed laser deposition (PLD) chamber and then transferred into a molecular beam epitaxy (MBE) chamber (Omicron Lab 10) with a vacuum $\sim$ 2 $\times$ 10$^{-10}$ mbar. High-purity Sb, Cr and Te were evaporated from Knudsen effusion cells. During growth of the Cr-doped Sb$_2$Te$_3$ film, the substrate was maintained at $\sim$ 240$^\circ$C. The flux ratio of $\frac{Te}{(Sb+Cr)}$ was set to be greater than 10 to prevent Te deficiency in the samples. Following the growth, the magnetic TI films were annealed at $\sim$ 240$^\circ$C for 30 minutes to improve the crystal quality before being cooled down to room temperature. Prior to the removal from the MBE chamber, a 10 nm thick Te layer was deposited at room temperature on top of the magnetic TI films for PNR measurements. See more details in our prior study [30].\\ 
\indent \textit{Polarized neutron reflectometry (PNR) measurement}: The samples were measured at 5 K in a saturating field to establish a base line. The magnetic field was then reduced to near remanence for a second measurement. The sample was subsequently warmed above the Néel temperature of Cr$_2$O$_3$ which showed no interface exchange coupling to allow us to better constrain the model fitting parameters. In all cases, the measurements were done from 0.01-0.15 1/\r{A} to distinguish the exchange at the interface and the bulk. 
\begin{figure}
  \centering
  \includegraphics[width=0.5\textwidth]{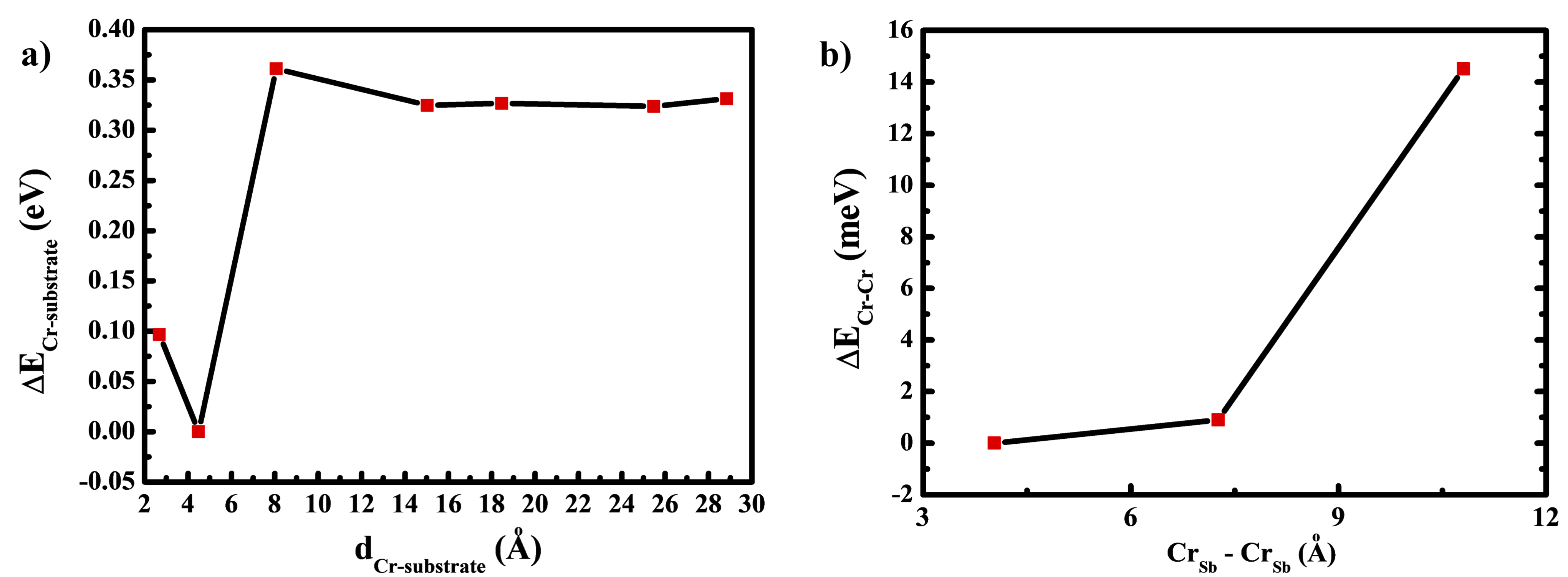}
  \caption {Thermodynamic properties of Cr dopants in the heterostructure. a) Formation energy of a single Cr dopant substituting Sb as a function of distance from the Cr$_2$O$_3$ substrate, with respect to the lowest formation energy. The formation energy is defined as E$_f$ = E$_{doped}$ - E$_{pristine}$ - n$\mu_{Cr}$ + m$\mu_{Sb}$, where E$_{doped}$ is the total energy of the Cr doped heterostructure, E$_{pristine}$ is the total energy of the pristine heterostructure, $\mu_{Cr}$ is the chemical potential of bulk Cr and $\mu_{Sb}$ is the chemical potential of Sb, and \textit{n} and \textit{m} are the numbers of Cr added and Sb atoms removed from the heterostructure. b) Total energy difference of 2 Cr dopants as a function of Cr-Cr separation distance.}
\end{figure}
\section{\label{sec:level3}Results and Discussions}
Before considering the effect of Cr dopants in the TI layer, we first consider the interfacial effect between the TIs (Bi$_2$Se$_3$ and Sb$_2$Te$_3$) that showed small lattice mismatch with Cr$_2$O$_3$. In the case of the Bi$_2$Se$_3$ interface, the TI thin film exhibits strong chemical bonding with the antiferromagnetic substrate as demonstrated by the small distance between the TI layer and Cr$_2$O$_3$ layers as illustrated in Table 1. The relaxed crystal structure of Bi$_2$Se$_3$/Cr$_2$O$_3$ showing the interfacial bonding is shown in Fig. S1. Due to the strong interaction between Bi$_2$Se$_3$ and Cr$_2$O$_3$, the electronic ground state of this heterostructure is half-metallic where the spin-up channel is metallic [Fig. S2b]. When the SOC is turned on, the heterostructure becomes full metallic [Fig. S2c]. As such this system cannot host a QAH phase. On the other hand, in the Sb$_2$Te$_3$/Cr$_2$O$_3$ heterostructure the TI layer shows no such chemical bonding with the Cr$_2$O$_3$ layer. This is also reflected in the larger interfacial distance between Sb$_2$Te$_3$ and the AFM substrate, compared to Bi$_2$Se$_3$, as shown in Table 1. Consequently, the Sb$_2$Te$_3$ TI potentially represents a higher quality interface with magnetic materials such as Cr$_2$O$_3$. Notice, that due to an A-type AFM ordering, the top-layer of the AFM insulator remains uncompensated [41], and can effectively induce spin-polarization in the TI layer due to the proximity coupling effect. However, since there is weak coupling between the TI and the AFM layers, the induced moments on the interfacial Te/Sb atoms were negligible ($\textless$ 0.01 $\mu B$/atom). To further increase the magnetic exchange interaction between the TI layer and the AFM substrate, we investigate doping the TI-layer with magnetic dopants.\\
\indent Among a wide range of possible magnetic dopants, we find Cr atoms that substitutionally dope the Sb-site to be the most favourable and study the electronic, magnetic and topological properties of the Cr-doped Sb$_2$Te$_3$/Cr$_2$O$_3$ heterostructure. To assess dopant segregation to surfaces/interfaces, we compute the total energy of the substitutional Cr-dopant across the TI film.  As shown in Fig. 2a, Cr-dopant has lowest formation energy at the interface with the AFM insulator. This suggests that Cr-dopants are likely to form at the TI/AFM-substrate interface. To see if the dopants would cluster, we next compare the total energy of a Cr-Cr dimer at the interface, as a function of distance (Fig. 2b).  We find that Cr-dopants have a tendency to cluster at the interface.  Further, this does not appear to introduce any unfavourable structural distortions or chemical bonding. This opens the possibility of increasing Cr-concentrations at the TI/AFM interface, achieving modulation doping.  As mentioned earlier, such modulation doping would result in a more homogeneous exchange gap.\\
\indent We next explore the nature of the magnetic exchange interaction of Cr-dopants in a cluster, as well as with the AFM substrate. When one Cr dopant is introduced in the TI layer at the interface, the magnetic interaction favors an antiferromagnetic exchange with the uncompensated top magnetic layer in the Cr$_2$O$_3$ substrate [Fig. 3a], with an exchange energy of $\sim$ 60 meV. This magnetic exchange coupling with Cr$_2$O$_3$ decreases quickly as the dopant moves farther away. To induce a QAHE, it is important that the magnetic coupling in the TI layer prefers the out-of-plane direction, thus we also calculated the magnetic anisotropy (MAE) by comparing the out-of-plane magnetic direction and the in-plane direction of the Cr dopants to determine the most favourable magnetic axis. According to Fig. 3a, at the interface the Cr-dopants strongly prefer to orient along the out-of-plane direction, while the Cr-dopants farther away from the interface, that are weakly coupled to the AFM substrate, fluctuate between an in-plane and an out-of-plane direction. This further suggests that Cr-dopants when at the interface will lead to a higher temperature QAHE, if their clusters are ferromagnetically coupled. \\
\begin{figure}[t]
  \centering
  \includegraphics[width=0.5\textwidth]{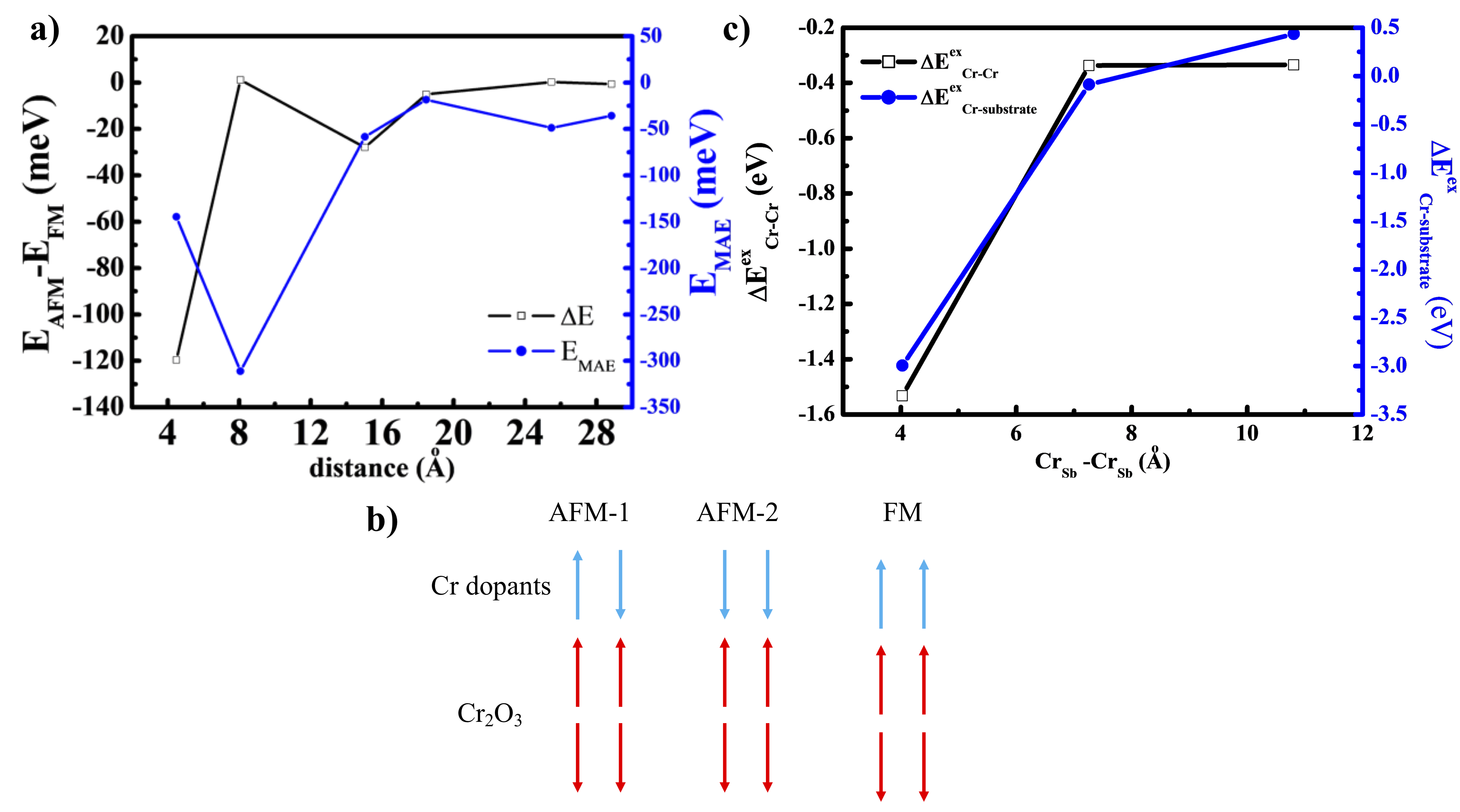}
  \caption {Cr dopant-dopant and dopant-substrate magnetic exchange interaction  in Cr doped Sb$_2$Te$_3$ and Cr$_2$O$_3$ heterostructure. a) Total energy difference of a single Cr-dopant in the TI-layer which is ferromagnetically or  antiferromagnetically coupled to the Cr$_2$O$_3$ substrate. A lower energy for the AFM arrangement suggests that Cr dopants at the interface are antiferromagnetically coupled to the top layer of Cr$_2$O$_3$ substrate.  The magnetic anisotropy ($E_{MAE}$) of the Cr dopant in Sb$_2$Te$_3$ is also plotted as a function of distance from the Cr$_2$O$_3$ substrate. b) Different magnetic configurations considered between the two Cr dopants in Sb$_2$Te$_3$ interfacial layer and the Cr$_2$O$_3$ substrate.  c) We plot $\Delta E_{Cr-Cr}^{ex} = E_{AFM-2} - E_{AFM-1}$ and $\Delta E_{Cr-substrate}^{ex} = E_{AFM-2} - E_{FM}$ as a function of Cr-Cr separation distance. This suggests that Cr-dopants prefer to cluster at the interface TI-layer.} 
\end{figure}
\begin{figure}[t]
  \centering
  \includegraphics[width=0.5\textwidth]{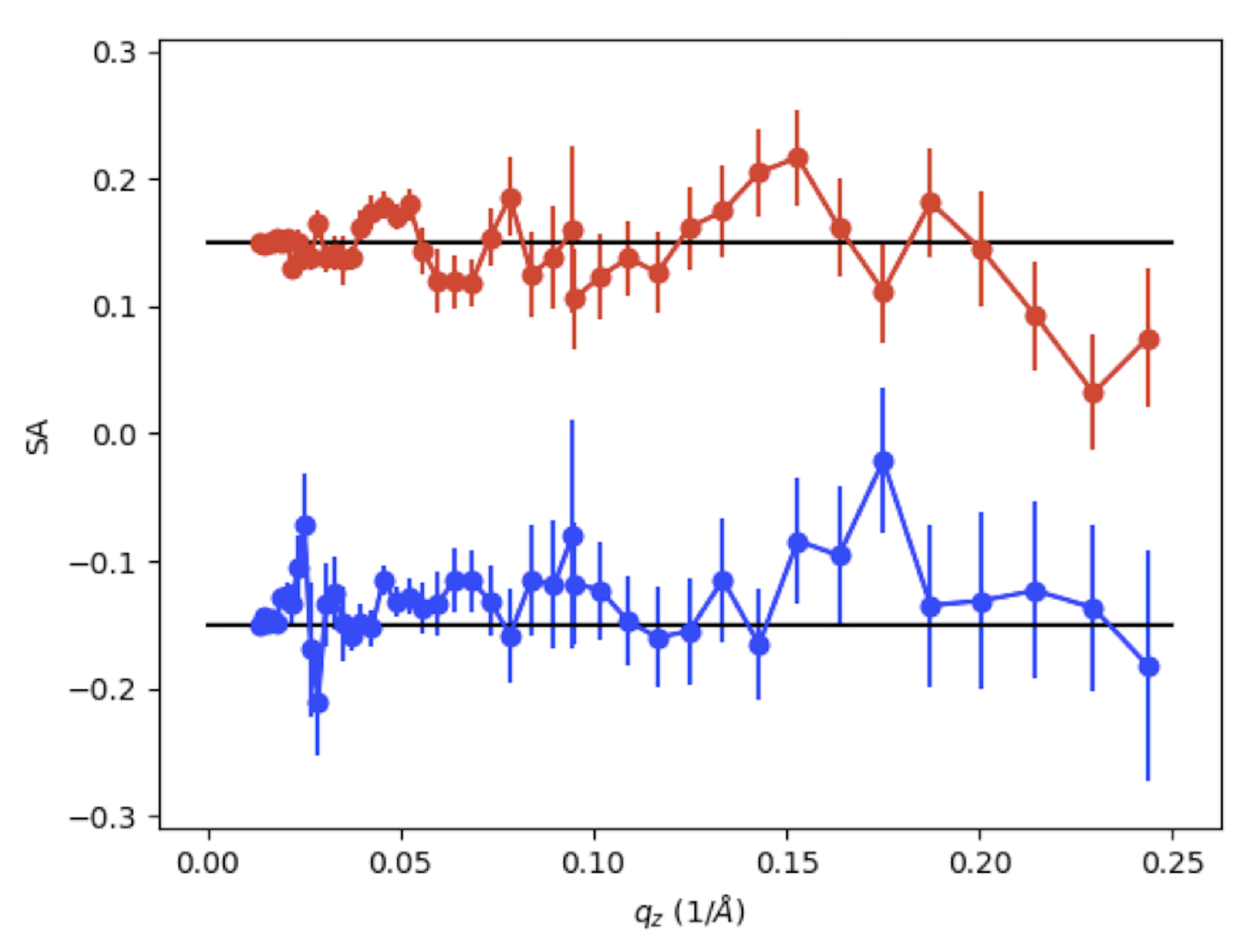}
  \caption {Polarized neutron measurement of four quintuple Sb$_{1.8}$Cr$_{0.2}$Te$_3$/Cr$_2$O$_3$, showing the spin asymmetry (SA=$(R^+ - R^-)/(R^+ + R^-)$) for the field cooled (Bottom) and zero field cooled (Top) cases. Here the data have been artificially shifted and individual “zero” lines drawn to allow easy comparison.} 
\end{figure}
\indent In addition to the coupling between the magnetic dopant and the antiferromagnetic substrate, to facilitate the QAH phase, the magnetic dopants need to couple ferromagnetically in the TI layer to induce a bulk magnetization in the TI-film, thereby breaking time-reversal symmetry. To consider the Cr-Cr magnetic interaction in the TI layer, we consider three different magnetic coupling between the two Cr dopants and the Cr$_2$O$_3$ layer as shown in Fig. 3b at different distances between the Cr dopants: i) (AFM-1) AFM coupling between the two Cr dopants and AFM coupling with the Cr$_2$O$_3$ layer, ii) (AFM-2) FM coupling between two Cr dopants and AFM coupling with the Cr$_2$O$_3$ layer, and iii) (FM) FM coupling between the two Cr dopants and FM coupling with the Cr$_2$O$_3$ layer. It should be noted that in our calculations we always kept one substituting Cr at the interface while varying the position of the other Cr dopants. According to Fig. 3c, the AFM-2 structure had the lowest energy.  When the Cr-dopants are clustered at the interface, they couple antiferromagnetically to the substrate, while remaining ferromagnetically coupled to each other. Previous theoretical study of Cr doped Sb$_2$Te$_3$  [35] has suggested that the FM coupling can survive even under a large separation of Cr dopants as d$_{Cr_{Sb}-Cr_{Sb}}$ $\approx$ 12 \r{A}. As a result, our results suggest this long-range ferromagnetism is unaffected in heterostructuring process with Cr$_2$O$_3$, which is also  in good agreement with the recent observation of an enhanced ferromagnetism in the Cr doped Sb$_2$Te$_3$ layer when interfacing with Cr$_2$O$_3$ [26]. Furthermore, the preferred AFM coupling between Cr dopants at the interface and the Cr$_2$O$_3$ layer also explains the observation of the positive exchange bias between Cr doped Sb$_2$Te$_3$ and the Cr$_2$O$_3$ substrate [30]. \\
\indent In addition to our DFT modeling of the magnetic exchange interaction at the interface, the magnetism in our 4QL Sb$_{1.8}$Cr$_{0.2}$Te$_3$/Cr$_2$O$_3$ heterostructure was also investigated experimentally using polarized neutron reflectometry (PNR). This technique is depth sensitive, thus allowing us to probe the magnetic fluctuation at different heights inside the doped TI thin film. As shown in Fig. 4, the spin asymmetry (SA) shows two different magnetic signatures for the field-cooled and zero field-cooled measurements showing the different magnetic layers spatially separated in the doped TI thin film along the z-axis. Following a field cool measurement,  an oscillation is observed in the SA immediately after the total reflection region (Fig. 4 blue curve) that indicates the presence of ferromagnetism throughout the bulk of the film. In the zero-field cooled case (Fig. 4 red curve), clear oscillations appear with a maximum near 0.05 1/\r{A} and again near 0.15 1/\r{A} indicating the presence of magnetic layers near the interface. Consequently, these results confirm our theoretical prediction of a stronger magnetic coupling at the interface between the doped TI and the antiferromagnetic Cr$_2$O$_3$ substrate, possibly due to dopant clustering. \\
\indent Transition metal doping with Cr has been predicted to open a gap on the Dirac surface of Sb$_2$Te$_3$ thin film up to a concentration of 10\% while maintaining a bulk insulating state [42-44]. To understand what other effects are introduced by the Cr$_2$O$_3$ surface on the magnetically doped-TI,  we calculated the electronic structures for different doping concentrations of Cr in the heterostructure with and without the Cr$_2$O$_3$ layer (Fig. 4-top-row(without) and 4-bottom-row(with)]. 
As shown in Fig. 4a, due to the small compressive strain and the hybridization effect between the top and bottom surfaces a significant inverted energy gap of 20.7 meV is observed in strained undoped free-standing 4 quintuple layer Sb$_2$Te$_3$ [Fig. 4a]. In the presence of Cr$_2$O$_3$, the energy gap at the $\Gamma$ point in the Sb$_2$Te$_3$ layer significantly reduces due to band-bending effects which also breaks the inversion symmetry of the free-standing TI-film. With the introduction of Cr in free-standing Sb$_2$Te$_3$, a metallic ground state is observed with increasing concentrations of Cr. In addition, Cr-dopants also induce a Zeeman gap which further increases the size of the inverted band gap at the $\Gamma$ point from 20.7 meV to 81.3 meV. The introduction of Cr at the surface of the free-standing Sb$_2$Te$_3$ also breaks inversion symmetry which induces pronounced asymetric Rashba splitting around the $\Gamma$ point as shown in Figs. 4b and 4c. This asymmetric Rashba splitting would also lead to inhomogeneous gaps. The metallic ground state can be attributed to the compressive strain which has been known to narrow the band bulk structure of Cr doped Sb$_2$Te$_3$ [45]. \\
\indent On the other hand, an insulating ground state is recovered in the heterostructure configuration between Cr doped Sb$_2$Te$_3$ and Cr$_2$O$_3$ [Figs. 4e and 4f] with a maximum inverted band gap of 67.5 meV. Most importantly, the Fermi level is inside the band gap. Furthermore, the direct gap at the $\Gamma$ point is shown to increase monotonically with increasing Cr-dopant concentration without shifting the Fermi-level i.e. keeping the Fermi-level in the middle of the band gap. Interestingly, as the Cr doping concentration increases at the interface the band structure reveals an increased contribution of a Rashba-like spin-orbit coupling to the band gap, in addition to the Zeeman splitting. This is due to an increasing contribution of inversion symmetry breaking with a higher concentration of defects. To interrogate the importance of Cr-dopants at the interface, we also looked at the nature of the gap as the Cr-dopant moves away from the interface. As shown in Fig. S4 the band structure undergoes a gap closing and opening transition with increasing Cr dopant distance from the interface. This is the signature of a topological transition with the non-trivial phase only emerging when Cr-dopants are at the interface, suggesting that likely Cr-dopants at surfaces/interfaces of TI-films induce non-trivial topology. Recent experimental studies have also suggested that the magnetic dopants at surfaces can exhibit different exchange coupling to the bulk, with the possibility of increasing the magnetization via modulation doping, thus potentially increasing the band gap [46]. Our results strongly support the idea of modulation doping in TI heterostructures with AFM-substrates to induce a large inverted band gap. \\
\begin{figure}[t]
  \centering
  \includegraphics[width=0.5\textwidth]{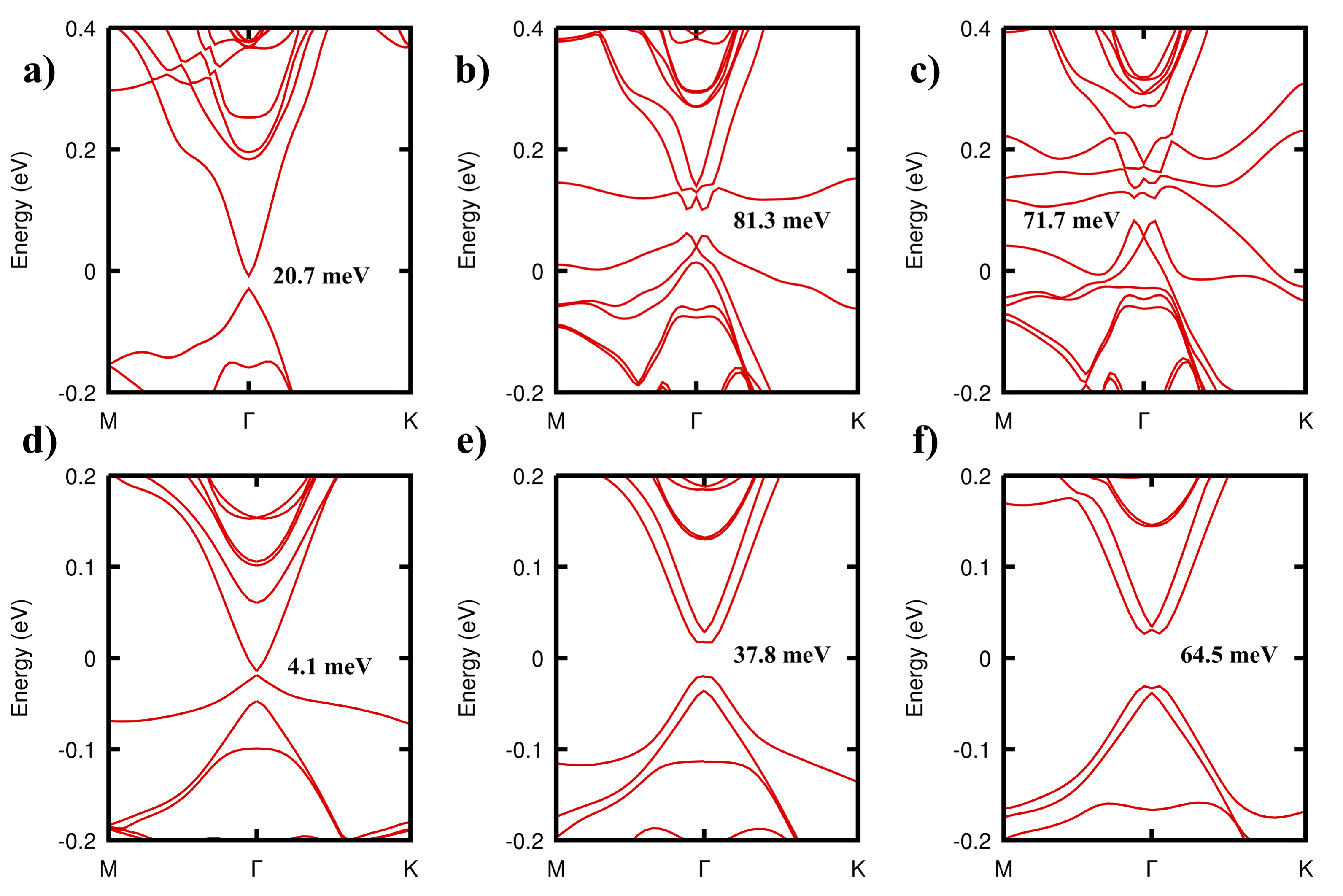}
  \caption {Electronic band structures of Sb$_2$Te$_3$ and Sb$_2$Te$_3$/Cr$_2$O$_3$ heterostructure. a), b) and c) show the band-structure around the Fermi-level without the Cr$_2$O$_3$ substrate for pristine, 1 Cr-dopant and 2 Cr-dopants at the interface TI-layer, respectively. d), e) and f) show the band-structure around the Fermi-level with the Cr$_2$O$_3$ substrate for the same three cases. The difference between the top and bottom panels clearly suggests non-trivial role of the Cr$_2$O$_3$ substrate on the electronic properties of the Cr-doped TI thin-film.} 
\end{figure}
\indent Band offset has been known to play an important role in pinning the Fermi level of Dirac points in TI/magnetic material heterostructure [47]. Based on our results outlined in Fig. 4, it is suggested that the Fermi level is pinned inside the band gap regardless of the doping concentration. Thus, to reveal the role of band offset in the heterostructure configuration, we plot the layer resolved density of states of Cr doped Sb$_2$Te$_3$/Cr$_2$O$_3$. As shown in Fig. S3, regardless of the doping concentration, the contributions from the TI's orbitals are always located inside the band gap of Cr$_2$O$_3$, thus indicating a signature of a type-I band offset. As a result, rather than inducing the magnetization on the TI layer as previous magnetic material/TI heterostructure Cr$_2$O$_3$ modifies the band structure of the doped Sb$_2$Te$_3$ via a type-I band offset to pin the Fermi level inside the bulk gap. \\
\begin{figure}[t]
  \centering
  \includegraphics[width=0.5\textwidth]{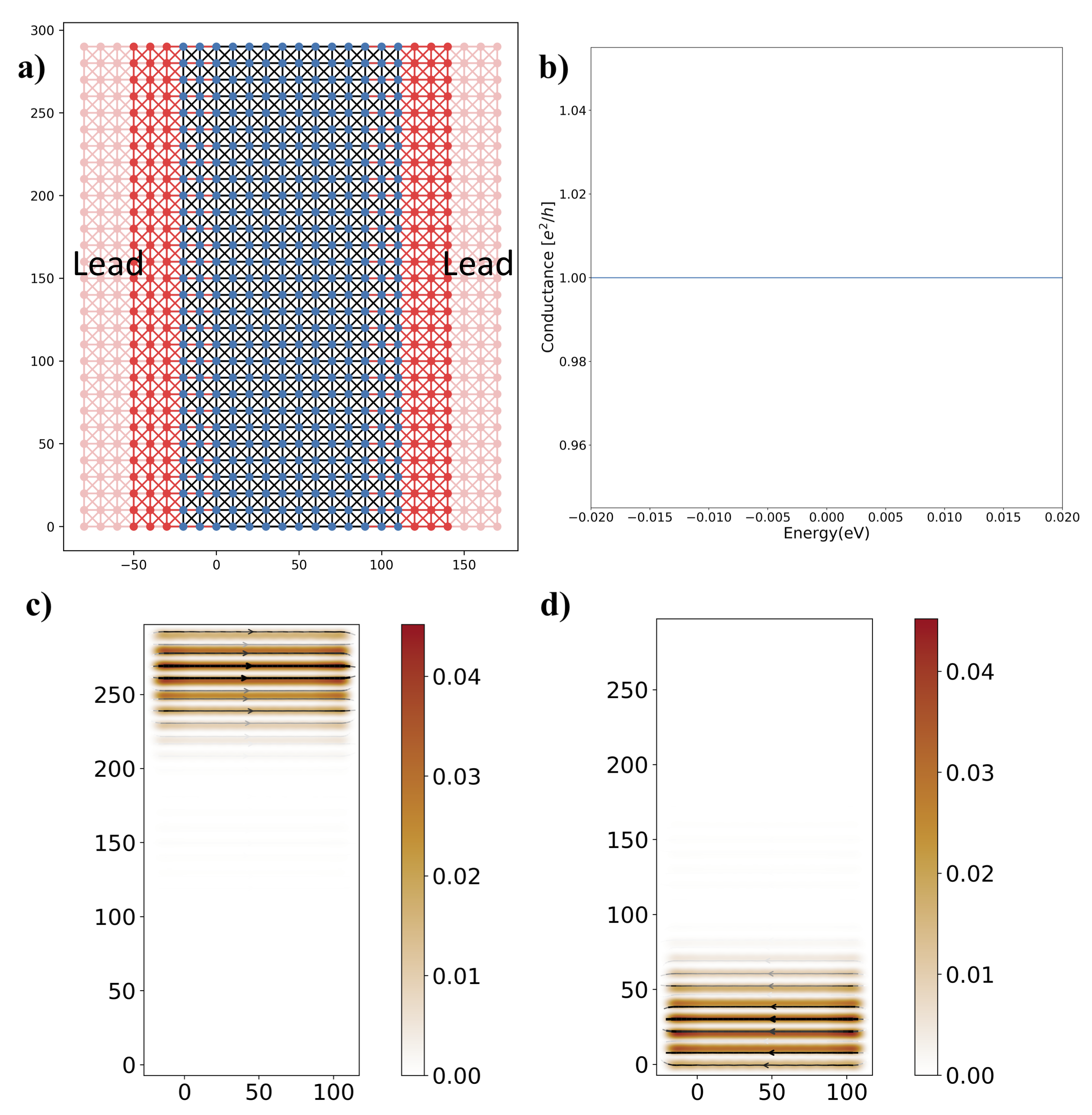}
  \caption{Transport  obtained by discretizing the {\sl k.p} Hamiltonian on a square lattice using parameters from table S1 for our heterostructure system containing a low concentration (i.e. 1 Cr-dopant in the supercell) of dopant at the interface. a) 2D square lattice with dimension 100L X 300W used for the transport calculation. b) Conductance measurement for the system showing a quantized value ($e^2/h$) indicating C = -1 expected for a QAH phase. c) Current direction corresponding to the top edge. d) Current direction for the bottom edge. Here the parameters were divided by an effective lattice of 12.584 \r{A} so that all the fitting parameters in the kwant program have the same unit of eV.}
\end{figure}
\indent Since the Cr doped Sb$_2$Te$_3$ represents an ideal interface with Cr$_2$O$_3$, we characterize the topological nature of the heterostructure at the two interfacial doping concentrations, using a 4 band ${\sl k.p}$ model [14,28] as described below by considering the states from both the top and bottom surfaces with different basis such as $\vert t \uparrow\rangle$, $\vert t \downarrow\rangle$, $\vert b \uparrow\rangle$, $\vert b \downarrow\rangle$. \\
$H_{total}$ = $H_{surface}$ + $H_{Zeeman}$ + $H_{interface}$ \\
$H_{surface}$ =\scalebox{0.9}{$\begin{pmatrix}
  A(k_x^2+k_y^2) & iv_{\alpha}k_- & M_k & 0\\ 
  -iv_{\alpha}k_+ & A(k_x^2+k_y^2) & 0 & M_k\\
  M_k & 0 & A(k_x^2+k_y^2) & -iv_{\alpha}k_-\\
  0 & M_k & iv_{\alpha}k_+ & A(k_x^2+k_y^2)\\
\end{pmatrix}$} \\
where \(M_k = M-B(k_x^2+k_y^2)\) is the hybridization between the top and bottom surfaces in the different spin direction, \(v_{\alpha} = v_F[1+\alpha(k_x^2+k_y^2)]\) is the second order correction to the Fermi velocity \(v_F\) [48], and \(k_{\pm} = k_x \pm ik_y \) is the Dirac dispersion.\\
$H_{Zeeman}$ = $\begin{pmatrix}
  \Delta & 0 & 0 & 0\\ 
  0 & -\Delta & 0 & 0\\
  0 & 0 & \Delta & 0\\
  0 & 0 & 0 & -\Delta\\
  \end{pmatrix}$ \\
where \(\Delta\) is the strength of Zeeman coupling originated from the magnetic dopant.\\ $H_{interface}$ = $\begin{pmatrix}
  V & 0 & 0 & 0\\ 
  0 & V & 0 & 0\\
  0 & 0 & -V & 0\\
  0 & 0 & 0 & -V\\
  \end{pmatrix}$ \\
where \(V\) is potential due to the inversion symmetry breaking (i.e. band-bending term).\\
\indent We fit the band structures from the DFT data of the configurations consisting one Cr-dopant and two-atom Cr-cluster at the interface [Fig S5] to the model Hamiltonian \(H_{total}\) around the $\Gamma$ point. The fitting parameters are presented in table S1. Based on the fitting parameters, the Chern numbers from the system were calculated to be -1 by using the Z2pack [49] which tracks the evolution of the Wannier charge center (WCC) as shown in Fig. S5. Further confirmation of the Chern insulating state was done by calculating the transport properties using the ${\sl kwant}$ package [50] as shown in Fig. 6. The conductance shows a quantized value of $e^2/h$ [Fig. 6b], and we also observe chiral edge currents [Fig. 6c and 6d], with opposite chirality at the top edge vs the edge at the interface with the AFM substrate. This is consistent with what we would expect in a QAH phase, characterized by a Chern number of C = -1. In addition, the results from ${\sl k.p}$ model suggest that the Cr doped Sb$_2$Te$_3$/Cr$_2$O$_3$ always exhibits a large Zeeman splitting which overcomes the effect of band bending effects that originates from inversion asymmetry in the TI layer due to the presence of Cr-dopants and an interface. Compared to previous studies of other interfaces  such as CrI$_3$/Bi$_2$Se$_3$ [28] the Zeeman coupling is shown to be an order of magnitude larger in this sytem, while the band gap of 64.5 meV  is comparable to the recent experimental discovery of a high temperature QAHE in MnBi$_2$Te$_4$/Bi$_2$Te$_3$ [51]. Consequently, the Cr doped Sb$_2$Te$_3$/Cr$_2$O$_3$ represents a promising heterostructure systems which can facilitate the realization of a high temperature QAHE. 
\section{\label{sec:level4}Conclusions}
\indent In conclusion, we have demonstrated that Cr doped Sb$_2$Te$_3$/Cr$_2$O$_3$ is a promising heterostructure to realize functional high temperature QAHE due to a combination of i) a large Zeeman energy from Cr dopants, that prefer to segregate to the interface and cluster, overcoming potential metallicity due to band bending effect; ii) Fermi level pinning inside the bulk gap due to a type-I band offset between the TI-films and the wide band gap AFM insulating substrate and iii) a nontrivial topological ground state protected by a Chern number of -1. Particularly, a low carrier concentration due to Fermi-level pinning in the center of the band gap makes it an ideal system to realize QAHE experimentally. Our study also opens up the possibility to utilize the AFM Cr$_2$O$_3$ substrate as a heterostructure with the recently discovered magnetic TI materials with ordered magnetic atoms, such as MnBi$_2$Te$_4$ [52,53] or to grow multilayers achieving higher-order Chern numbers [20]. Also, growing conventional {\sl s}-wave superconductors on top of this heterostructure should also induce topological superconductivity in these quantized edge-states at the interface. 
\section*{Acknowledgement}
A.D.P. was financially supported by the Oak Ridge National Laboratory’s Laboratory Directed Research and Development project (Project ID 7448, PI: P.G.). The DFT calculations used resources of the National Energy Research Scientific Computing Center (NERSC), a U.S. Department of Energy Office of Science User Facility operated under Contract No. DE-AC02-05CH11231, via the Center for Nanophase Materials Sciences, a US Department of Energy Office of Science User Facility and the Oak Ridge Leadership Computing Facility at the Oak Ridge National Laboratory, which is supported by the Office of Science of the U.S. A portion of this research used resources at the Spallation Neutron Source, a DOE Office of Science User Facility operated by the Oak Ridge National Laboratory. These measurement are associated with proposal number 22026. C.-Z.C are supported by the Gordon and Betty Moore Foundation’s EPiQS Initiative (Grant GBMF9063 to C.Z.C.) and the ARO Young Investigator Program Award (W911NF1810198). 
\section*{References}
\noindent{[1] C.-X. Liu, S.-C. Zhang, and X.-L. Qi, Annu. Rev. Condens. Matter Phys. 7, 301 (2016).} \\
\noindent{[2] K. He, Y. Wang, and Q.-K. Xue, Annu. Rev. Condens. Matter Phys. 9, 329 (2018).} \\
\noindent{[3] J. Wang, B. Lian, S.-C. Zhang, Phys. Scr. 2015, 014003 (2015).} \\
\noindent{[4] Q. L. He, L. Pan, A. L. Stern, E. C. Burks, X. Che, G. Yin, J. Wang, B. Lian, Q. Zhou, E. S. Choi, K. Murata, X. Kou, Z. Chen, T. Nie, Q. Shao, Y. Fan, S.-C. Zhang, K. Liu, J. Xia, K. L. Wang. Science. 357, 294 (2017).} \\
\noindent{[5] R. Yu, W. Zhang, H.-J. Zhang, S. C. Zhang, X. Dai, Z. Fang, Science 329, 61 (2010).}\\
\noindent{[6] C. X. Liu, X. L. Qi, X. Dai, Z. Fang, S. C. Zhang, Phys. Rev. Lett. 101, 146802 (2008).} \\
\noindent{[7] C.-Z. Chang, J. Zhang, X. Feng, J. Shen, Z. Zhang, M. Guo, K. Li, Y. Ou, P. Wei, L.-L. Wang, Z.-Q. Ji, Y. Feng, S. Ji, X. Chen, J. Jia, X. Dai, Z. Fang, S.-C. Zhang, K. He, Y. Wang, L. Lu, X.-C. Ma, Q.-K. Xue, Science 340, 167 (2013).}\\
\noindent{[8] C.-Z. Chang, W. Zhao, D. Y. Kim, H. Zhang, B. A. Assaf, D. Heiman, S.-C. Zhang, C. Liu, M. H. W. Chan, and J. S. Moodera, Nat Mater 14, 473 (2015).}\\
\noindent{[9] C.-Z. Chang, M. Li, J. Phys. Condens. Matter 28, 123002 (2016).}\\
\noindent{[10] C.-Z. Chang, J. Zhang, M. Liu, Z. Zhang, X. Feng, K. Li, L.-L. Wang, X. Chen, X. Dai, Z. Fang, X.-L. Qi, S.-C. Zhang, Y. Wang, K. He, X.-C. Ma, and Q.-K. Xue. Adv. Mater. 25, 1065 (2013).}\\
\noindent{[11] J. Zhang, C.-Z. Chang, P. Tang, Z. Zhang, X. Feng, K. Li, L.-l. Wang, X. Chen, C. Liu, W. Duan, K. He, Q.-K. Xue, X. Ma, Y. Wang, Science 339, 1582 (2013).}\\
\noindent{[12] Y. Feng, X. Feng, Y. Ou, J. Wang, C. Liu, L. Zhang, D. Zhao, G. Jiang, S.-C. Zhang, K. He, X. Ma, Q.-K. Xue, and Y. Wang, Phys. Rev. Lett. 115, 126801 (2015).}\\
\noindent{[13] X. Kou, L. Pan, J. Wang, Y. Fan, E. S. Choi, W.-L. Lee, T. Nie, K. Murata, Q. Shao, S.-C. Zhang, and K. L. Wang, Nat. Commun. 6, 8474 (2015).}\\
\noindent{[14] X. Feng  Y. Feng, J. Wang, Y. Ou, Z. Hao, C. Liu, Z. Zhang, L. Zhang, C. Lin, J. Liao, Y. Li, L.‐L. Wang, S.‐H. Ji, X. Chen, X. Ma, S.‐C. Zhang, Y. Wang, K. He,  Q.‐K. Xue, Adv Mater. 28, 6386 (2016).}\\
\noindent{[15] I. Lee, C. K. Kim, J. Lee, S. J. L. Billinge, R. Zhong, J. A. Schneeloch, T. Liu, T. Valla, J. M. Tranquada, G. Gu, and J. C. S. Davis, Proc. Natl. Acad. Sci. U.S.A. 112, 1316 (2015).}\\
\noindent{[16] C. Z. Chang, P. Z. Tang, Y. L. Wang, X. Feng, K. Li, Z. C. Zhang, Y. Y. Wang, L. L. Wang, X. Chen, C. X. Liu, W. H. Duan, K. He, X. C. Ma, and Q. K. Xue, Phys. Rev. Lett. 112, 056801 (2014).}\\
\noindent{[17] C. Z. Chang, P. Z. Tang, Y. L. Wang, X. Feng, K. Li, Z. C. Zhang, Y. Y. Wang, L. L. Wang, X. Chen, C. X. Liu, W. H. Duan, K. He, X. C. Ma, and Q. K. Xue, Phys. Rev. Lett. 112, 056801 (2014).}\\
\noindent{[18] M. Mogi, R. Yoshimi, A. Tsukazaki, K. Yasuda, Y. Kozuka, K. S. Takahashi, M. Kawasaki, Y. Tokura, Appl. Phys. Lett. 107, 182401 (2015).}\\
\noindent{[19] J. Jiang, D. Xiao, F. Wang, J. H. Shin, D. Andreoli, J. X. Zhang, R. Xiao, Y. F. Zhao, M. Kayyalha, L. Zhang, K. Wang, J. D. Zang, C. X. Liu, N. Samarth, M. H. W. Chan, and C. Z. Chang, Nat. Mater. 19, 732 (2020).}\\
\noindent{[20] Y. F. Zhao, R. Zhang, R. Mei, L. J. Zhou, H. Yi, Y. Q. Zhang, J. Yu, R. Xiao, K. Wang, N. Samarth, M. H. W. Chan, C. X. Liu, and C. Z. Chang, Nature 588, 419 (2020).}\\
\noindent{[21] F. Katmis, V. Lauter, F. S. Nogueira, B. A. Assaf, M. E. Jamer, P. Wei, B. Satpati, J. W. Freeland, I. Eremin, D. Heiman, P. Jarillo-Herrero, J. S. Moodera, Nature 533, 513 (2016).}\\
\noindent{[22] J. Kim, K.-W. Kim, H. Wang, J. Sinova, and R. Wu. Phys. Rev. Lett. 119, 027201 (2017).}\\
\noindent{[23] A. Kandala, A. Richardella, D. W. Rench, D. M. Zhang, T. C. Flanagan and N. Samarth, Appl. Phys. Lett. 103, 202409 (2013).}\\
\noindent{[24] W. M. Yang, S. Yang, Q. H. Zhang, Y. Xu, S. P. Shen, J. Liao, J. Teng, C. W. Nan, L. Gu, Y. Sun, K. H. Wu and Y. Q. Li, Appl. Phys. Lett. 105, 092411 (2014).}\\
\noindent{[25] L. D. Alegria, H. Ji, N. Yao, J. J. Clarke, R. J. Cava, and J. R. Petta, Appl. Phys. Lett. 105, 053512 (2014).}\\
\noindent{[26] M. Lang, M. Montazeri, M. C. Onbasli, X. Kou, Y. Fan, P. Upadhyaya, K. Yao, F. Liu, Y. Jiang, W. Jiang, K. L. Wong, G. Yu, J. Tang, T. Nie, L. He, R. N. Schwartz, Y. Wang, C. A. Ross and K. L. Wang, Nano Lett. 14, 3459 (2014).} \\
\noindent{[27] C. Tang, C.-Z. Chang, G. Zhao, Y. Liu, Z. Jiang, C.-X. Liu, M. R. McCartney, D. J. Smith, T. Chen, J. S. Moodera, and J. Shi, Sci. Adv. 3, 1700307 (2017).} \\
\noindent{[28] Y. Hou, J. Kim and R. Wu, Sci. Adv. 5, eaaw1874 (2019).}\\
\noindent{[29] A. Pham, P. Ganesh, arXiv:2003.05840 (2020).}\\
\noindent{[30] F. Wang, D. Xiao, W. Yuan, J. Jiang, Y.-F. Zhao, L. Zhang, Y. Yao, W. Liu, Z. Zhang, C. Liu, J. Shi, W. Han, M. H. W. Chan, N. Samarth, C.-Z. Chang, Nano Lett. 19, 2945 (2019).}\\
\noindent{[31] S. Zhu, D. Meng, G. Liang, G. Shi, P. Zhao, P. Cheng, Y. Li, X. Zhai, Y. Lu, L. Chen, and K. Wu, Nanoscale, 10, 10041 (2018).}\\
\noindent{[32] J. Li, C.B. Wilson, R. Cheng, M. Lohmann, M. Kavand, W. Yuan, M. Aldosary, N. Agladze, P. Wei, M. S. Sherwin, J. Shi, Nature 578, 70(2020).}\\
\noindent{[33] S.-Y. Jeong, J.-B. Lee, H. Na, T.-Y. Seong, Thin Solid Films 518, 4813 (2010).}\\
\noindent{[34] M. Lucht, M. Lerche, H.-C. Wille, Yu. V. Shvyd'ko, H. D. Rüter, E. Gerdau, and P. Becker, J. Appl. Cryst. 36, part 4, 1075 (2003).} \\
\noindent{[35] J.-M. Zhang, W. Ming, Z. Huang, G.-B. Liu, X. Kou, Y. Fan, K. L. Wang, and Y. Yao, Phys. Rev. B 88, 235131 (2013).}\\
\noindent{[36] A. Pham and P. Ganesh, Phys. Rev. B 100, 241110(R) (2019).}\\
\noindent{[37] S. M. Frolov, M. J. Manfra, J.D. Sau. Nat. Phys. 16, 718–724 (2020).}\\
\noindent{[38] J. P. Perdew, K. Burke, and M. Ernzerhof, Phys. Rev. Lett. 77, 3865 (1996).} \\
\noindent{[39] P. E. Blochl, Phys. Rev. B 50, 17953 (1994).} \\
\noindent{[40] S. L. Dudarev, G. A. Botton, S. Y. Savrasov, C. J. Humphreys, and A. P. Sutton, Phys. Rev. B 57, 1505 (1998).} \\
\noindent{[41] A. L. Wysocki, Siqi Shi, and K. D. Belashchenko, Phys. Rev. B 86, 165443 (2012).} \\
\noindent{[42] J. Kim, S.-H. Jhi, A. H. MacDonald, and R. Wu, Phys. Rev. B 96, 140410(R) (2017).}\\
\noindent{[43] M. F. Islam, C. M. Canali, A. Pertsova, A. Balatsky, S. K. Mahatha, C. Carbone, A. Barla, K. A. Kokh, O. E. Tereshchenko, E. Jiménez, N. B. Brookes, P. Gargiani, M. Valvidares, S. Schatz, T. R. F. Peixoto, H. Bentmann, F. Reinert, J. Jung, T. Bathon, K. Fauth, M. Bode, and P. Sessi, Phys. Rev. B 97, 155429 (2018).}\\
\noindent{[44] M. Ye, W. Li, S. Zhu, Y. Takeda, Y. Saitoh, J. Wang, H. Pan, M. Nurmamat, K. Sumida, F. Ji, Z. Liu, H. Yang, Z. Liu, D. Shen, A. Kimura, S. Qiao, and X. Xie, Nat. Commun. 6, 8913 (2015).}\\
\noindent{[45] Y. Ruan, Y. Yang, Y. Zhou, L. Huang, G. Xu, K. Zhong, Z. Huang, and J.-M. Zhang, J. Phys.: Condens. Matter 31, 385501 (2019).} \\
\noindent{[46] W. Liu, Y. Xu, L. He, G. van der Laan, R. Zhang, K. Wang, Sci. Adv. Vol. 5, eaav2088 (2019).}\\
\noindent{[47] T. V. Menshchikova, M. M. Otrokov, S. S. Tsirkin, D. A. Samorokov, V. V. Bebneva, A. Ernst, V. M. Kuznetsov, E. V. Chulkov. Nano Lett. 13, 6064 (2013).}\\
\noindent{[48] L. Fu, Phys. Rev. Lett. 103, 266801 (2009).}\\
\noindent{[49] D. Gresch, G. Autès, O. V. Yazyev, M. Troyer, D. Vanderbilt, B. A. Bernevig, and A. A. Soluyanov Phys. Rev. B 95, 075146 (2017).}\\
\noindent{[50] C. W. Groth, M. Wimmer, A. R. Akhmerov, X. Waintal, New J. Phys. 16, 063065 (2014).}\\
\noindent{[51] E. D. L. Rienks, S. Wimmer, J. Sánchez-Barriga, O. Caha, P. S. Mandal, J. Růžička, A. Ney, H. Steiner, V. V. Volobuev, H. Groiss, M. Albu, G. Kothleitner, J. Michalička, S. A. Khan, J. Minár, H. Ebert, G. Bauer, F. Freyse, A. Varykhalov, O. Rader, and G. Springholz, Nature 576, 423 (2019).} \\
\noindent{[52] W. Ko, M. Kolmer, J. Yan, A. D. Pham, M. Fu, F. Lüpke, S. Okamoto, Z. Gai, P. Ganesh, and A.-P. Li, Phys. Rev. B 102, 115402 (2020).} \\
\noindent{[53] F. Lüpke, A. D. Pham, Y.-F. Zhao, L.-J. Zhou, W. Lu, E. Briggs, J. Bernholc, M. Kolmer, W. Ko, C.-Z. Chang, P. Ganesh, A.-P. Li, arXiv:2101.08247 (2021).}\\

\end{document}